\documentclass{class_nature}
\usepackage[T1]{fontenc}
\usepackage{textcomp}
\usepackage{gensymb}
\usepackage{amsmath} 
\usepackage{color}
\usepackage{graphicx}
\usepackage{xfrac}
\usepackage{soul} 
\usepackage{lineno} 
\usepackage[english]{babel}
\usepackage{blindtext}
\usepackage{amssymb}
\usepackage{caption}
\usepackage{array}
\usepackage{makecell}
\usepackage{comment}
\usepackage{float}
\usepackage{hyperref}
\usepackage{url}
\usepackage{siunitx}

\hypersetup{
    colorlinks=false,
    linkcolor=blue,
    filecolor=magenta,      
    urlcolor=cyan,
    pdftitle={Overleaf Example},
    pdfpagemode=FullScreen,
    }

\graphicspath{ {./Figures/} }

\makeatletter
\DeclareRobustCommand\bfseriesitshape{%
  \not@math@alphabet\itshapebfseries\relax
  \fontseries\bfdefault
  \fontshape\itdefault
  \selectfont
}
\makeatother

\newcommand{\beginsupplement}{%
        \setcounter{table}{0}
        \renewcommand{\thetable}{S\arabic{table}}%
        \setcounter{figure}{0}
        \renewcommand{\thefigure}{S\arabic{figure}}%
     }

\title{multi-Stochastic Core Architecture for Scaling Probabilistic Ising Machines} 

\author{Chirag Garg$^{1,*}$, Pratik Brahma$^{1,*}$, Saavan Patel$^{1,*}$, Sayeef Salahuddin$^{1}$}

\begin{document}

\maketitle

\begin{affiliations}
 \item Department of Electrical Engineering and Computer Sciences, University of California, Berkeley, California 94720, USA \\
 $^{*}$ These authors contributed equally to this work
\end{affiliations}


\begin{abstract}

Ising Machines \cite{Patel2022LogicallyFactorization, supriyo2019integer, saavan2024pass, peter_coherent_ising_machine, moy2022coupled, efficientoptimization, chowdhury2023accelerated} offer vast potential to solve NP-hard combinatorial optimization problems \cite{simulated_annealing_kirkpatrick, mohseni2022ising, lucas_mapping} efficiently that are intractable to solve using conventional computing architecture. A lot of these optimization problems fall into statistical learnability and involve finding an optimal solution among many possible, near-analogous configurations, by searching in a non-convex energy landscape \cite{simulated_annealing_kirkpatrick}. In this context, the probabilistic Boltzmann machine architecture especially PASS – Parallel Asynchronous Stochastic Sampler \cite{saavan2024pass}, explores and models the complex probability landscape pertaining to all possible configurations and excels in finding the ground-state energy solution of these intractable problems. Additionally, the noise-based neuron architecture addresses the limitation of conventional annealing methods, which may get stuck around local minima. Here, we demonstrate a stochastic sampling approach based on Block Gibbs Sampling to integrate multiple asynchronous PASS chips (four in this work), enabling improved scalability.
Further, we demonstrate the scaling by mapping $\sim 800$ nodes Max-Cut problem integrating 256 nodes PASS accelerator manufactured in state-of-the-art 14 nm CMOS FinFET technology.  PASS-enabled system with Block Gibbs Sampling protocol shows $\sim 1000$ times speedup for Max-Cut optimization compared to state-of-the-art methods implemented on CPUs and GPUs. The general applicability of this approach is further illustrated by solving a quantum spin chain Transverse Ising system and accurately representing complex probability landscapes. Moreover, our results demonstrate the change in the scaling law to constant in the scaled-PASS accelerator as compared to exponential on GPUs enabling at least 4 orders of magnitude improvement in time-to-solution. Hence, the presented methodology enables the pathway for scaling of asynchronous brain-like dynamics systems that do not follow any clock for its operation.

\end{abstract}

Combinatorial Optimization problems are prevalent across numerous domains such as logistics \cite{logistics_ref, TSP_Dac_alberta}, finance \cite{arbitrage_iscas}, biology \cite{lidar_bio_application}, integrated-circuit design \cite{fpga_placement}, etc. These problems require the exploration over an exponentially large solution space which becomes computationally intractable for conventional computing solutions. Generally, such problems are classified as NP-hard, as no polynomial-time algorithms are known for solving them exactly on conventional computers. Simulated annealing is conventionally a common heuristic technique used to find the best or nearly best solutions of combinatorial optimization \cite{simulated_annealing_kirkpatrick}. Alternatively, hardware accelerators based on Ising models are leveraged to overcome the computational bottleneck posed by these optimization problems \cite{mohseni2022ising}. Further, the decision-making process employed by Ising models resembles that of many complex optimization problems, enabling their transformation into Ising Hamiltonians or cost functions \cite{lucas_mapping}.

The Ising Hamiltonian, in its mathematical form, abstracts real-world hardware dynamics, leading to a spectrum of performance and efficiency outcomes across different Ising-based hardware architectures. These architectures can be divided into two broad classes. In the first class, architectures employ discrete-time system dynamics similar to software-based annealing methods in which neuron/spin-update happens sequentially \cite{reconfigurable_ising_sa, Yamaoka2016AAnnealing}. As a result, the performance of these Ising machines is limited by the number of spin-updates, which rise exponentially with the problem's size. In contrast, the second class features natural coupling dynamics that inherently allow parallelized spin-updates and enable them to efficiently solve complex optimization problems at scale \cite{peter_coherent_ising_machine, chris_kim_osc_vlsi, moy2022coupled, supriyo2019integer, saavan2024pass}. Most of the Ising solvers in this class \cite{peter_coherent_ising_machine, chris_kim_osc_vlsi, moy2022coupled} only aim to find the ground state solution for a given input problem. Given that hard optimization problems are largely statistical \cite{simulated_annealing_kirkpatrick}, probabilistic architectures \cite{supriyo2019integer, saavan2024pass, Patel2022LogicallyFactorization} offer wider applicability and can comprehensively represent the solution space for a given problem.

\begin{figure*}[h!]
\begin{centering}
\includegraphics[width=\linewidth]{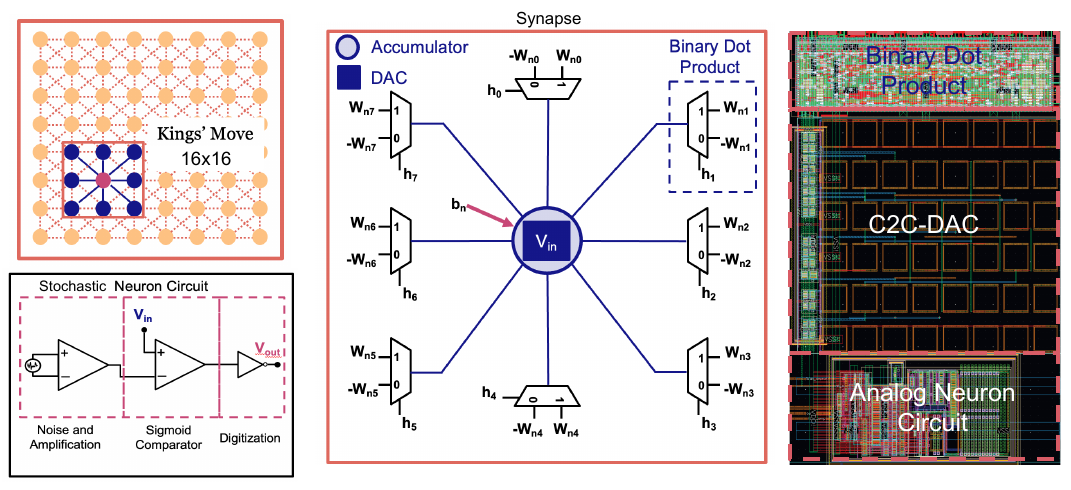}
\par\end{centering}
\caption{\label{fig:full_circuit}\textbf{Overview of single-core PASS implementation. }  \\
Top-left:  Top-level neuron diagram where 256 neurons are arranged in a grid and connected in King's move fashion. Bottom-left: Stochastic neuron circuit which consists of a Noise source with an amplifier, sigmoid comparator as an activation to neuron, and digitization stage to produce stochastic binary samples. Center: Synapse calculation circuit with multiplexers used for the multiplication of weights and neighboring neuron signals, and digital-to-analog conversion of accumulated multiplications. Right: Layout of analog neuron and synapse circuit. 
}
\end{figure*}

Probabilistic architectures comprise stochastic neurons/spins interconnected via a synapse network (Extended Data Fig. \ref{fig:full_circuit}) and employ energy-based principles derived from Boltzmann distributions to guide the state of the system. It is represented as 
\begin{gather}
    p(s) = \frac{1}{Z} e^{-\beta E(s)} \label{eq:boltzmann1}\\ 
    E(s) = \sum_i \sum_j J_{ij} s_i s_j + \sum_i h_i s_i, \ \  s \in \{-1, 1\} \label{eq:boltzmann2}
\end{gather}
where $\beta$ is introduced as an inverse temperature modulating probabilities associated with energy minima; Z, partition function; $J_{ij}$, the interaction between stochastic neuron $s_{i}$ and $s_{j}$; and $h_{i}$, bias term for neuron $s_{i}$. These architectures are deeply linked with Restricted Boltzmann machines-based neural networks \cite{Ackley1985AMachines} that excel at capturing complex patterns and relationships within data through unsupervised learning. In this work, we leverage a fully integrated CMOS-based PASS stochastic core \cite{saavan2024pass} that exploits its massive parallelism and asynchronous nature resulting in orders of speedup over conventional computation solutions. Further, this core is a statistical sampling system that employs true randomness to replicate the annealing phenomenon and tackle a wide class of combinatorial optimization problems \cite{lucas_mapping}. It sets itself apart from other Ising-model accelerator IC (integrated circuits) implementations due to its capability to address machine learning \cite{Hinton2002TrainingDivergence, Onizawa2020In-HardwareLearning}, neural decision-making \cite{Sridhar2021TheCollectivesb} and quantum physics problems \cite{quantum_problem_kerem}.

Highly parallelized and clockless Ising systems provide substantial acceleration and energy savings compared to their traditional counterparts \cite{saavan2024pass}. However, the conventional approach of integrating smaller chips (chiplets) into a large-scale system using high-speed interconnects is not directly applicable to these systems. It is because these systems do not rely on a clock for their spin-update operation. Scaling via chiplets integration becomes important to address the escalating computational needs while navigating challenges such as die reticle constraints and yield limits in advanced process nodes \cite{intel_highspeed_interconnects_dassharma2024high, multi_core_packaging}. To overcome this challenge, our approach of stochastic sampling provides a pathway. 

In this work, we present a Block Gibbs sampling approach that integrates four PASS accelerators, enabling improved scalability and achieving an efficiency of 18.67 binary-TOPS/W at 0.6 V operation. However, configuring a network over multiple cores/chips in a way that not only allows going to a larger network but also provides a scaling in the time to solution is not trivial. The general applicability of this approach on any probabilistic Ising machine is illustrated by mapping randomly generated Max-Cut problem instances and still achieving $\sim 1000$ times time-to-solution (TTS) acceleration. Further, the efficacy of this sampling-based scaling approach is demonstrated by mapping quantum Monte-Carlo models on PASS and precisely modeling the probability distribution to a granularity of one hundred millionths ($10^{-8}$). We particularly use a lattice of qubits described by a transverse field Ising model to establish this result and report at least 4 orders of magnitude performance improvement over Nvidia A100 GPUs. Therefore, the Block Gibbs sampling approach offers a promising pathway for scaling of parallel, asynchronous (or clockless), and energy-efficient probabilistic hardware to meet growing computational needs. 

\section*{Scaling Approach}

PASS stochastic core adheres to Boltzmann machine dynamics that probabilistically sample neurons in the system given by their pairwise interactions (equations \ref{eq:boltzmann1} and \ref{eq:boltzmann2}). A wide range of NP-hard optimization problems can be reduced to the Ising Hamiltonian form \cite{lucas_mapping} in equation \ref{eq:boltzmann2}  that resembles a graph network with node values representing the state and edges corresponding to interaction weights (Fig. \ref{fig:mapping_scheme}a). This graph network is then mapped \cite{embedding} onto the locally connected synapse network of PASS core (Fig. \ref{fig:mapping_scheme}b) to sample ground state probability distribution (Fig. \ref{fig:mapping_scheme}d). However, problem size that can be solved is limited by the number of neurons in a single core. Our proposed approach, termed multi-Stochastic Cores Architecture (m-SCA), integrates multiple cores (Fig. \ref{fig:mapping_scheme}c), expanding machine sizes to handle larger problems and enhancing computational throughput, rather than scaling up a single core. 

The practical applicability of an Ising machine is determined by its scalability and speed performance. In this context, we present a pathway that underscores the key principle of stochastic sampling to integrate many stochastic cores/chips for scaling problem sizes. The approach involves partitioning the neurons of the mapped optimization problem into multiple cores/blocks, specifically four in this case (Fig. \ref{fig:mapping_scheme}c). Each core in this multi-core system comprises stochastic neurons of a PASS accelerator chip which is independently sampled once during the burn-in phase. During the sampling phase, we use the concept of conditional probability and sample each block of the network, given the states of its neighboring blocks from the previous iteration. A full probability distribution can then be established based on these conditional probabilities of each block. This technique is called Block Gibbs Sampling. Further, the conditional probability is set by clamping the adjacent edge neurons based on the stored states of neighboring blocks from the previous iteration (Fig. \ref{fig:mapping_scheme}c). Ultimately, the whole neuron system is sampled, and a ground-state energy solution is achieved (Fig. \ref{fig:mapping_scheme}d). It is also worth noting that block-wise sampling does not mean that when one block is sampled, the other blocks are simply idle. Since the conditional probability is defined with respect to the states of the neighboring blocks in the previous iteration, continuous sampling can go on resulting in enhanced computational throughput.

\section*{Max-Cut Optimization Problem}

\begin{figure*}
\begin{centering}
\includegraphics[width=\linewidth]{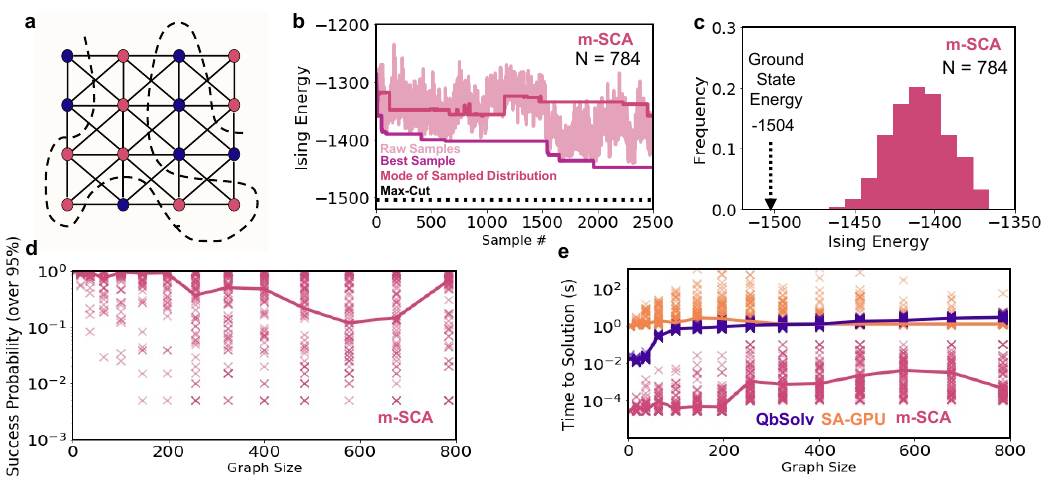}
\par\end{centering}
\caption{\label{fig:maxcut} \textbf{Max-Cut Optimization Problem} \\ 
{\textbf{a}} Pictorial representation of randomly generated King's move connectivity graph instances for finding Max-Cut solution. 
{\textbf{b}} Evolution of Max-Cut Ising Energy with samples in the final iteration of the sampling phase and demonstration of the two methods to interpret samples from m-SCA.
{\textbf{c}} Cut values distribution obtained from 200 independent experiments and compared against ground truth given by qbsolv.
{\textbf{d}} Success probabilities of random instances of Max-Cut problem up to 784 nodes.
{\textbf{e}} Time to Solution to achieve 99 $\%$ success probability and comparison with efficient simulated annealing-based GPU implementation.
}
\end{figure*}

In this work, we investigate the computational accuracy of the proposed multi-Stochastic Cores Architecture (m-SCA) by solving Max-Cut problem instances on it. Max-Cut is an NP-hard optimization problem \cite{Karp1972_optimization_ref} that seeks to split the vertices of a graph into two distinct subsets which is called a cut and maximize the sum of weighted edges comprising that cut (Fig. \ref{fig:maxcut}a). This problem reveals the general applicability of the proposed scaling approach as it exactly resembles the Ising model form. To validate the performance, we benchmark the Max-Cut results of the proposed architecture (m-SCA) against the simulated annealing (SA-GPU) implemented with parallelized Gibbs sampling on the Nvidia A100 Tensor Core GPU and classical tabu search algorithm on Intel Xeon Gold 6330 processor. 

\begin{figure*}
\begin{centering}
\includegraphics[width=0.88\linewidth]{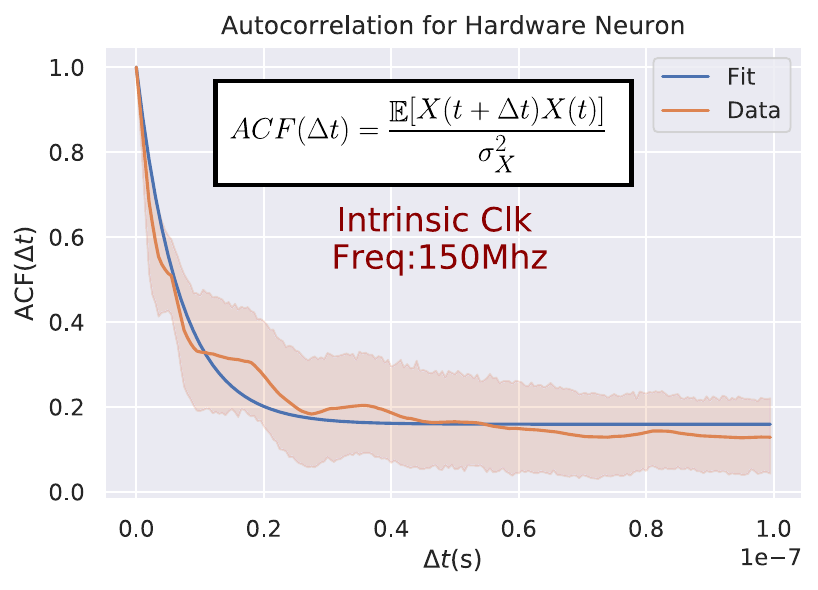}
\par\end{centering}
\caption{\label{fig:acf}\textbf{Autocorrelation fit of m-SCA system}  \\
Autocorrelation time extracted from the sampled data of PASS core. The exponential decaying fit to the data defines the rate of generating independent stochastic samples. Therefore, it sets the system dynamic and determines the speed of the internal spin update clock (150 MHz).}
\end{figure*}

We program m-SCA comprising four PASS accelerators with 100 randomly generated Max-Cut problems for each graph size, weights $\{-1, 1\}$ and node count ranging from 4 to 784. These sparsely-connected Max-Cut problems on King's move graphs (Fig. \ref{fig:maxcut}a) are directly mapped onto m-SCA. In a $16\times16$ neuron PASS core, a $14\times14$  neuron cluster is utilized for programming the main problem instance, and the edge neurons around it are clamped to set the conditional probability based on the states of the adjacent core in m-SCA. The largest configuration of m-SCA cluster in this work supports 784 node problems (6.2k connections) with a computational speed reaching 940.8 GMACs per second (1 GMAC = $10^{9}$ Multiply and accumulate operations) at 150 MHz Poisson update clock (Extended Data Fig. \ref{fig:acf}) and 14.7 GigaSamples/s at 18.75 MHz sampling clock. Hence, the capacity of m-SCA cluster with 4 PASS cores has scaled up by a factor of 3.06 in terms of total neurons (3.37 in connections) and 3.37 times in computational (MAC) throughput compared to the single-core PASS implementation \cite{saavan2024pass} (256 spins and 279 GMACs per second). Further, the 4 PASS cores system consumes 50.4 mW at an operating voltage of 0.6 V, achieving a high energy efficiency of 18.67 binary-TOPS/W, with one operation counted per MAC.

Using the proposed m-SCA, we perform 4 sequential computation steps in the sampling phase of Block Gibbs sampling (Fig. \ref{fig:mapping_scheme}c) and sample every core 510 times during each computation. Fig. \ref{fig:maxcut}b shows the raw samples produced during a sequential computation step while solving a 784-node Max-Cut problem. The stochastically produced raw samples tend to move amongst and towards the highest probability/lowest energy state. These samples can be analyzed in two different ways to find the ground-state solution. The first method involves the mode estimation of collected samples in each sequential step and uses the best one to determine the highest probability state. It is related to the mixing time of Markov Chain \cite{Bremaud1999_markovchain}. In the second method, the lowest energy state encountered among the sampled states is used as an estimate of the highest probability state. This method is linked to the hitting time of Markov Chain \cite{Bremaud1999_markovchain}. Here, we employ the second method because it effectively follows the dynamics of energy distribution toward the ground-state solution. Since the conditional probability is not set frequently, the first method of mode estimation tends to be heavily affected by the lower probability states captured from m-SCA during the first few steps of the sampling phase. We use the same procedure to solve each Max-Cut problem instance and run experiments 200 times to measure the Ising energy distribution (Fig. \ref{fig:maxcut}c) and calculate success probability. Fig. \ref{fig:maxcut}c depicts the energy distribution for 784-node Max-Cut problem and benchmarks its convergence near the best solution found by the classical tabu search algorithm (qbsolv) \cite{qbsolv}.

For performance benchmarking, we adopt the success probability ($p_s$) metric and define success for the experiments as obtaining a cut value greater than 95 $\%$ of the best solution found by the qbsolv framework. The median success probability across the 100 randomly generated problems for each Max-Cut graph sizes is plotted in Fig. \ref{fig:maxcut}d as a solid line. The crosses indicate a spread in success probabilities, primarily attributed to the varying difficulty levels of solving different random instances within each size category. This success probability metric captures both performance and energy consumption. The higher value implies fewer attempts to solve a particular problem to reach ground state energy solution, leading to shorter TTS and lower energy consumption. Fig. \ref{fig:maxcut}e shows the total computation time required to obtain solution ($>$ 95 $\%$ accuracy) with a 99 $\%$ success probability.
It is calculated by 
\begin{equation}
    \label{eq:TTS}
    TTS = T_{comp}*\frac{log(1-0.99)}{log(1-p_{s})} \\ 
\end{equation}
where $T_{comp}$ is the total computation taken to sample the states of all sequential steps during the sampling phase of an experiment. It confirms the solvability on m-SCA with a significant acceleration of $\sim 1000$ times compared to simulated annealing on GPU and qbsolv on CPU. However, the performance acceleration is traded off to achieve high computational throughput via m-SCA visible for problem sizes greater than 196 nodes. Thus, the proposed scaling architecture m-SCA demonstrates the computing advantage for solving large-scale optimization problems.

\section*{Quantum 1D Transverse Ising}

In the current era, where general-purpose quantum computers are yet to be realized, Quantum Monte Carlo (QMC) continues to be the standard method for exploring quantum many-body systems \cite{quantum_monte_carlo_science,chowdhury2023accelerated}. These techniques are pivotal for delving into various quantum phenomena efficiently as they sample equilibrium statistics of basis-states to capture the wavefunction that define these systems. In particular, the probability distribution derived from these samples helps measure the system in those basis-states. In this context, we emulate the 1D quantum chain system following stoquastic Transverse Field Ising Hamiltonian \cite{transverse_ising_hamiltonian, quantum_problem_kerem} onto the proposed m-SCA. It leverages the highly parallelized and efficient sampling to accelerate QMC sampling compared to conventional hardware systems. Further, the solvability of this problem is a testament to the applicability of statistical Block Gibbs sampling based scaling approach in accurately modeling the probability distribution, extending even to the complexities of quantum wavefunctions.

\begin{figure*}
\begin{centering}
\includegraphics[width=0.88\linewidth]{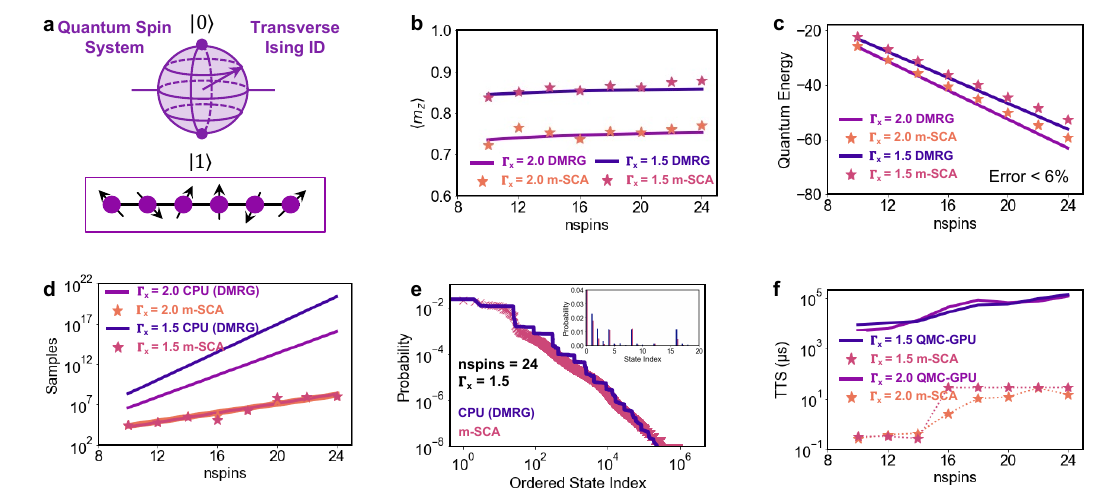}
\par\end{centering}
\caption{\label{fig:quantum} \textbf{Quantum 1D Transverse Ising} \\
\textbf{a} Quantum spin chain which is trotterized to map onto classical m-SCA system.
Evaluation of m-SCA samples and {\textbf{b}} benchmarking of the order parameter (average magnetization) and {\textbf{c}} quantum energy against theoretical DMRG calculations. 
\textbf{d} Total samples taken out of m-SCA to achieve the quantum energy error within 6$\%$. It is compared against the samples that DMRG calculations require to precisely define the probabilities of basis states. 
\textbf{e} The probability distribution of ordered states, that define the quantum spin chain Hamiltonian, given by m-SCA samples matches with theoretical DMRG calculations. 
\textbf{f} Time to Solution (TTS) scaling on m-SCA and GPU for quantum chain size up to 24 spins. m-SCA follows constant scaling law as compared to exponential scaling in GPUs.
}

\end{figure*}

The Transverse Ising Hamiltonian in one dimension (1D) to analyze N spin quantum chain (Fig. \ref{fig:quantum}a) is given as  
\begin{equation}
    \label{eq:transverse_ising}
    H = -\left(\sum_{i=1}^{N} J_{i,i+1} \sigma_i^z \sigma_{i+1}^{z} + \Gamma_x \sum_{i=1}^{N} \sigma_i^x + \Gamma_z \sum_{i=1}^{N} \sigma_i^z \right).
\end{equation}
where $\sigma_i$ represents the spin operator for $i_{th}$ spin in the chain, $J_{i,i+1}$; interaction strength between neighboring spins, $\Gamma_z$; symmetry-breaking magnetic field along z-direction, and $\Gamma_x$; transverse magnetic field applied along x-direction to introduce quantum fluctuation. This 1D-spin quantum Hamiltonian is converted into to its 2D-spin classical equivalent using Suzuki-Trotter decomposition \cite{suzzuki_trotter}. It results in a classical spin lattice with N spins and M replicas interacting with each other. We set the spin coupling interaction $J_{i,i+1}$ = 1 corresponding to ferromagnetic phase in the spin-chain system \cite{Humeniuk_2020_long_range_order} and magnetic field $\Gamma_z$ = 1. Under this condition, the system exhibits long-range order aligning all spins along the z-axis given the quantum fluctuations introduced by $\Gamma_x$ are not strong enough to disrupt the ferromagnetic order. The phenomenon is defined by the order parameter $\langle m_z \rangle$, average magnetization defined by the operator $\sum \sigma_j^z/N$. With these settings, N spin chain system is mapped onto m-SCA system to examine the order state of the system under transverse field ($\Gamma_x$ = 1.5 $\&$ 2.0). Fig. \ref{fig:quantum}b clearly shows that the larger transverse field in x-direction ($\Gamma_x$) weakens the spin order in z-direction, and hence $\langle m_z \rangle$. We also report the quantum energy of the system illustrating the physics defined by the quantum Hamiltonian, i.e., higher transverse field $\Gamma_x$ and increased interaction terms in larger chain system lower the energy of quantum system (Fig. \ref{fig:quantum}c).

We benchmark $\langle m_z \rangle$ and quantum energy using density-matrix renormalization group method (DMRG) \cite{SCHOLLWOCK201196_dmrg} implemented on CPU. DMRG is a crucial tool to study the statics and dynamics of strongly-correlated quantum systems. For the comparison, QMC is implemented on m-SCA and run only once during sampling-phase to approximate the system properties. In Fig. \ref{fig:quantum}b, the $\langle m_z \rangle$ results sampled from m-SCA show a good agreement with theoretical calculations from DMRG. Further, the quantum energy of the system from sampled states approximately matches with DMRG within 6$\%$ error range. This error is an artifact of exponentially growing samples requirement for larger spin system to accurately measure it with sampled basis states (Fig. \ref{fig:quantum}d). These exponentially large samples lead to computational challenges in calculating the quantum energy limiting the study to 24 spins with 24 replicas. Despite this complexity, the samples from m-SCA precisely model the probability distribution of ordered states with probability as low as $10^{-8}$ (Fig. \ref{fig:quantum}e) that will lead to an accurate wave-function. This indicates that the quantum energy error is an accumulated result of low-probability ($<10^{-8}$) states.

\begin{figure*}
\begin{centering}
\includegraphics[width=0.88\linewidth]{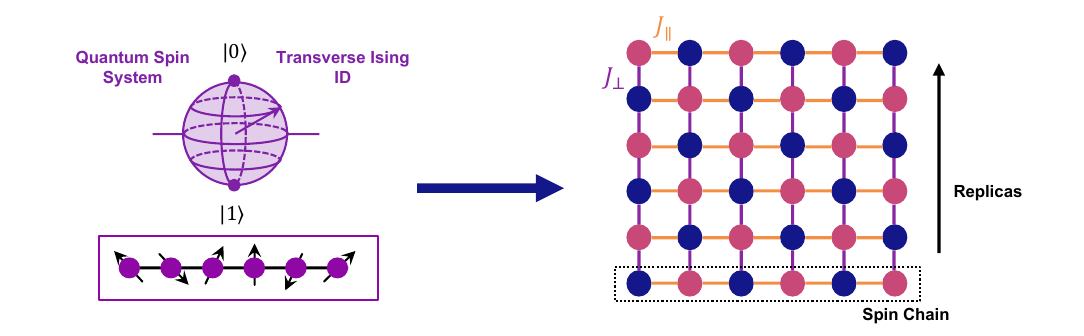}
\par\end{centering}
\caption{\label{fig:transverese_maping}\textbf{Trotterization of 1D quantum spin chain}  \\
It depicts the decomposition of a 1-dimensional quantum spin chain into a 2-dimensional lattice of spins.}
\end{figure*}

The quantum energy convergence with theoretical DMRG calculation defines the performance of m-SCA. TTS is defined as the time-step for which the accumulated samples up to that point yield 6$\%$ quantum energy error compared to theoretical calculations. We also implement the QMC on the Nvidia A100 Tensor Core GPU (QMC-GPU) to establish the performance acceleration achieved by the m-SCA system. Traditionally, single-flip QMC sampling is employed which updates only one-spin in a clock cycle, therefore, often takes too long to converge to the ground-state solution. Instead, we leverage a massively parallelized QMC scheme \cite{chowdhury2023accelerated} on GPU that graph-color all the neurons/spins on classically decomposed lattice into two sets (Extended Data Fig. \ref{fig:transverese_maping}) and update all of them in just two cycles. Fig. \ref{fig:quantum}f demonstrates the time-scaling behavior of optimized QMC on a GPU compared with m-SCA hardware. QMC-GPU exhibits exponential scaling with spin size, whereas m-SCA maintains approximately constant scaling.
It is attributed to parallelized and asynchronous neuron updates resulting in almost constant sample time, while highly parallelized QMC suffers from limited parallelism and the synchronous nature of GPUs.  A sudden change in the TTS curve for m-SCA around 14 spins results from the transition from single-core m-SCA computation to four-core m-SCA. Finally, Fig. \ref{fig:quantum}f confirms that m-SCA achieves at least 4 orders of magnitude acceleration of QMC methods compared to state-of-the-art GPUs. Thus, quantum simulation on classical m-SCA systems has emerged as a valuable arena for the accelerated investigation of various quantum phenomena in the future.

\section*{Conclusion}

Probabilistic, highly parallelized, and clockless Ising accelerators have been widely explored in literature \cite{supriyo2019integer, saavan2024pass, chowdhury2023accelerated} and known to tackle a broad range of computationally hard problems compared to their counterparts. We have presented multi-Stochastic Cores Architecture (m-SCA) to scale their processing capabilities by a factor of 3.37 compared to single-core architecture and achieve a significantly high efficiency of 18.67 binary-TOPS/W. The core concept of Block Gibbs sampling guides the m-SCA implementation. To generalize its applicability on the hard problems reducible to Ising Hamiltonian, m-SCA solves the Max-Cut graph up to 784 nodes, achieves $\sim 1000$ times speed-up relative to GPU, and consumes around 64.3 $\mu$W average power per neuron operation including synapse calculations. We have also mapped the quantum spin chain following Transverse Ising 1D Hamiltonian and shown the potential of m-SCA to capture the equilibrium statistical distribution of states with a probability precision of $10^{-8}$. m-SCA captures the physics of the spin-chain quantum system by correctly determining order parameter $\langle m_z \rangle$ and quantum energy with an accuracy above 94$\%$. It also significantly accelerates the QMC achieving at least 4 orders of magnitude speed-up compared to GPUs. This significant acceleration is attributed to asynchronous updates due to uncorrelated samples produced at a frequency of 150 MHz and ultra-fast sampling of 18.75 Megasamples per neuron per sec. Thus, m-SCA provides a significant acceleration of NP-hard classical and quantum problems over GPUs in a restricted power budget. Due to these performance advantages, this architecture presents itself as a prominent candidate to scale the computational and sampling throughput of a stochastic system.

\begin{figure*}

\begin{centering}
\includegraphics[width=\linewidth]{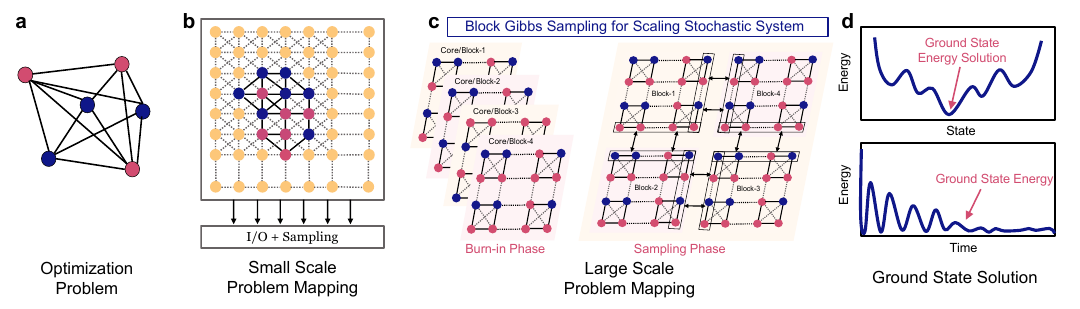}
\par\end{centering}
\caption{\label{fig:mapping_scheme} \textbf{Overview of multi-Stochastic Cores Architecture (m-SCA).} \\ 
{\textbf{a}} Optimization problem represented as a graph. 
{\textbf{b}} Mapping of the problem nodes and their pairwise interaction onto Parallel Asynchronous Stochastic Sampler (PASS) integrated circuit core. 
{\textbf{c}} Block Gibbs sampling approach implemented over multi-PASS cores to scale the system capabilities. Large-scale problems are mapped onto m-SCA available neurons and partitioned into blocks. These blocks are first sampled independently during the burn-in phase. However, in the sampling phase, neurons are conditionally sampled given the states of neighboring blocks to establish the probability distribution across all neurons in m-SCA. 
{\textbf{d}} These states are sampled over the complex non-convex energy landscape of the optimization problem and ultimately converge to the ground state energy solution.
}

\end{figure*}

\clearpage
\newpage
\begin{methods}

\section*{PASS Stochastic Core}
PASS core comprises 256 stochastic mixed-signal neurons operating at 0.6 V, locally connected synapses in King’s move configuration, and a built-in sampler (Extended Data Fig. \ref{fig:full_circuit}). Each stochastic neuron uses shot noise locally generated by a reverse bias diode and is sufficiently amplified to serve as input to the sigmoid comparator. The sigmoid comparator acts as a probabilistic activation function on input $V_{in}$ coming from neuron synapse and compares it with the amplified noise to generate a digitized stochastic signal. Since the core does not use any clock for computation, this digitized stochastic signal serves as an internal clock for asynchronous neuron updates in the system. The autocorrelation decay of the Poisson process determines the speed of the internal clock (150 MHz, Extended Data Fig. \ref{fig:acf}). 

The synapse essentially functions as a multiply-accumulate unit that operates on the output of neighboring neurons and interaction weights. In this binarized stochastic system, the multiplication is simplified to a digital multiplexer. These accumulated multiplications are converted into analog $V_{in}$ via a Digital-to-Analog Converter (DAC) which asynchronously updates the neurons. 

The data from neurons is sampled via an external clock (300 MHz) from the FPGA host system. This clock samples each neuron every k (number of rows in the 2D neuron array of the chip) cycles. Therefore, the effective sampling rate for each neuron is 18.75 MHz. Detailed insights are available in reference \cite{saavan2024pass}. 

\section*{Max-Cut Problem Setup}
Max-Cut is an optimization problem and is represented in Ising Hamiltonian for a graph G with edges E as
\begin{equation}
    \label{eq:maxcut}
       H(s) =  \sum_{\{i,j\} \in E(G)} J_{ij} s_i s_j , \ \  s \in \{-1, 1\}
\end{equation}

This Hamiltonian aims to partition the graph with interaction weight $J_{ij}$ into two sets $V^{+}$ ($s_i = 1$) and $V^{-}$ ($s_i = -1$) such that the sum of edge weights between these two sets is maximized. It is derived as follows using equation \ref{eq:maxcut}:

\begin{equation}
    \label{eq:maxcut_deriv_1}
       H(s) =  \sum_{\{i,j\} \in E (V^+)} J_{ij}  + \sum_{\{i,j\} \in E (V^-)} J_{ij} - \sum_{\{i,j\} \in \delta(V)} J_{ij}
\end{equation}
where $E (V^+)$ represents set of edges connecting nodes in $V^{+}$, $E (V^-)$ connecting  $V^{-}$, and $\delta(V)$ interconnecting $V^{+}$ and $V^{-}$. 

\begin{equation}
    \label{eq:maxcut_deriv_2}
       H(s) =  \sum_{\{i,j\} \in E(G)} J_{ij} - 2\sum_{\{i,j\} \in \delta(V)} J_{ij}
\end{equation}

Hence, equation \ref{eq:maxcut_deriv_2} concludes that minimizing the Ising Hamiltonian $H(s)$ implies finding the maximum cut $\sum_{\{i,j\} \in \delta(V)} J_{ij}$. For benchmarking the Max-Cut on m-SCA, we leverage equation \ref{eq:maxcut} to calculate the Ising Hamiltonian energy for the sample cut values and derive the success probability metric. 

\section*{Transverse Ising 1D Setup}
Transverse Ising Hamiltonian (equation \ref{eq:transverse_ising}) of the quantum spin chain is converted into its classical equivalent using Suzuki-Trotter decomposition as follows:
\begin{equation}
    \label{eq:transverse_ising_classical}
    H_{classical} = - \left( \lim_{M \to \infty} \sum_{k=1}^{M} \sum_{i=1}^{N} (J_{\parallel})_{i, i+1} \, s_{i,k} s_{i+1,k} + \gamma_z s_{i,k} + J_{\perp} \, s_{i,k} s_{i,k+1} \right)
\end{equation}

where $(J_{\parallel})_{i, j}$ = $J_{i, j}/M$, $\gamma_z$ = $\Gamma_z/M$, $J_{\perp}=-1/(2\beta)\log\tanh(\beta\Gamma_x/M)$, and  $s \in \{-1, 1\}$. These weights and biases are real-valued, whereas m-SCA handles 8-bit integer precision. Therefore, the constant weight multiplication factor and bias addition is applied to all neurons to handle the precision issue. In equation \ref{eq:transverse_ising_classical}, the equivalent classical Hamiltonian is represented as a 2-dimensional lattice (Extended Data Fig. \ref{fig:transverese_maping}) with N spins and M replicas. This representation exactly replicates the quantum spin chain system under the limit of infinite replicas. However, finite replicas are chosen keeping the error in permissible limits. \cite{trotter_error.aaa4170}. This finite MxN 2-dimensional lattice system forms the subgraph of King's move connectivity graph, therefore, is directly mapped onto m-SCA system and partitioned into blocks for sampling (Fig. \ref{fig:mapping_scheme}c). However, at higher values of $\Gamma_x$ (e.g., $\Gamma_x=2$), the perpendicular interaction becomes negligible, resulting in a system of $M$ effectively independent spin chains. Each chain consists of $N$ spins and can therefore be directly mapped onto a block of $N$ contiguous neurons. The $M$ replicas can then be distributed across the blocks in the m-SCA architecture.

\section*{Quantum Energy Calculation}

Quantum state can be expressed as 
\begin{equation}
    \label{eq:quantum state}
        |\psi\rangle \ = \sum_{x}  |x\rangle \langle x|\psi \rangle = \sum_{x}  \psi(x)|x\rangle
\end{equation}

The expected value over quantum Hamiltonian H which is denoted as quantum energy of the system E.
\begin{gather}
    \langle H \rangle = \frac{\langle \psi|H|\psi \rangle}{\langle \psi|\psi \rangle} =   \frac{\sum_{x} \langle \psi|x \rangle \langle x|H|\psi \rangle}{\sum_{x} \langle \psi|x \rangle \langle x|\psi \rangle} = E \frac{\sum_{x} \langle \psi|x \rangle \langle x|\psi \rangle}{\sum_{x} \langle \psi|x \rangle \langle x|\psi \rangle} (\therefore H|\psi \rangle = E |\psi \rangle) \label{eq:quantum energy1}\\
    E = \frac{\sum_{x} |\langle \psi|x \rangle|^2 \frac{\langle x|H|\psi \rangle}{\langle x|\psi \rangle} }{\sum_{x} |\langle \psi|x \rangle |^2} =  \frac{\sum_{x} | \psi(x) |^2 \frac{\langle x|H|\psi \rangle}{\langle x|\psi \rangle} }{\sum_{x}  |\psi(x)|^2} \label{eq:quantum energy2}
\end{gather}

Using equation \ref{eq:quantum energy2}, quantum energy\cite{quantum_energy_book} is calculated using the sampled states from m-SCA. 

\end{methods}


\newpage
\begin{addendum}

\item [Acknowledgements] 
This work is supported by MURI project  from Department of Defense and Office of Naval Research.

\item [Author Contributions] 
C.G. formulated the m-SCA scheme. C.G. and P.B. performed the CPU/GPU benchmarks. S.P. designed the PASS accelerator and C.G. performed m-SCA experiments on it. C.G. and S.S. co-wrote the manuscript. S.S. supervised the research.
All authors contributed to discussions and commented on the manuscript. 

\item [Competing Interests] 
The authors declare that they have no competing financial interests.

\item [Correspondence] 
Correspondence and requests for materials can be addressed to either C.G. \\ (chirag$\_$garg@berkeley.edu) or S.S. (sayeef@berkeley.edu). 

\item [Code and Data Availability]
Code and Data will be made available on reasonable request by emailing C.G. or S.S.

\end{addendum}


\clearpage

\clearpage

\clearpage

\clearpage


\newpage
\bibliographystyle{naturemag}
\bibliography{references.bib}


\onecolumn
\clearpage
\section*{Extended Data}
\beginsupplement


\end{document}